# Continual Model of Medium II: a Universal Adaptive Algorithm for the Triangulation of a Smooth Molecular Surface

O. Yu. Kupervasser[a,*] and N. E. Wanner[b]

[a]TRANSIST VIDEO Ltd., Skolkovo resident

[b]All-Russia Research Institute of Veterinary Sanitary, Hygiene, and Ecology, Russian Academy of Agricultural Sciences, Moscow

*e-mail: olegkup@yahoo.com

**Abstract**—In this paper, we represent an algorithm describing the original and universal principles of the triangulation of a smooth molecular solvent excluding surface (SES) obtained by primary and secondary rolling and a solvent accessible surface (SAS) that is derivative of SES. These surfaces are the boundaries between a molecule and a solvent.

The originality of our work consists in developing a universal and adaptive triangulation algorithm. The universality of the triangulation algorithm consists in that it is suitable not only for a rolling surface composed of toroidal and spherical fragments, but also for any smooth surface, including any level surface. The adaptivity of this algorithm consists in that the size of a triangulation mesh may be varied depending on its location, thus taking into account even small, but smooth surface features and preventing the "jump" onto close, but non-adjacent surface domains, and excluding the "cut-off" of narrow necks and channels. This is attained by reducing the triangulation grid step to a value less than the two principal curvature radii of a surface and also near an "active center" representing an area, within which non-adjacent surface domains approach each other at a small distance.

The obtained triangulated surface may be used for demonstration purposes in molecular editors (the algorithm itself is suitable for the triangulation of any smooth surface, e.g., a level surface) and also for the calculation of the solvation energy in continual solvent models.

## 1. Introduction

In this paper, we describe a universal and adaptive algorithm for the triangulation of a surface and the formation of surface elements. This algorithm will further be used to calculate the solvation energy gradient from atomic shifts on the basis of determining the gradients of the parameters of surface elements. An important and distinctive feature of our algorithm in comparison with may other algorithms for the triangulation of a molecular surface consists in that the boundaries of triangular surface meshes lying near the boundaries of the toroidal and spherical segments of a molecular surface do not generally belong to the boundary between these segments. This considerably simplifies the calculation of the gradients of the parameters of such boundary surface elements.

Let us give a brief review of the works devoted to the triangulation of a molecular surface.

To construct the surface of a substrate molecule, we surround all its atoms with spheres of van der Waals radius [1–3]. To obtain a smooth surface required to provide the convergence of the method, the surface is usually subjected to the primary and secondary "rolling" with spheres. The surface is then triangulated, i.e., partitioned into small triangles. These triangles are further used to form surface elements.

There exist the two types of smooth molecule surrounding surfaces, the formation of which was detailed in the previous works [4–9], namely:

(1) SAS, a solvent accessible surface, is formed by the centers of solvent molecules tangent to a substrate molecule; and

(2) SES, a solvent excluded surface. The volume occupied by a solvent lies *outside* the volume enveloped by this surface. The substrate itself lies completely *inside* this volume.

The following steps are the triangulation of a surface and the construction of surface elements on the basis of these triangles. There exists a wide class of algorithms designed for the triangulation of molecular surfaces. Most of these algorithms are non-universal and tied to a certain surface formation algorithm and type of formed surface segments [10–19]. The universal marching cubes (MC) algorithm and its improved dual contouring (DC) analogue do not take into account the individual properties of a smooth surface, thus complicating the construction of a uniform grid [20–22].

In the given work, we represent a universal triangulation algorithm that is suitable for the triangulation of not only a rolling surface, but also any level surface. It belongs to the class of advancing front technique (AFT) algorithms with direct surface meshing [23–28]. The other used names are the continuation method through the predictor-corrector [29–32], the mesh generation using processing sequences [33], or the paving method [34].

A triangulation grid is generated via the sequential addition of triangles onto a surface using the algorithm of the projection of an arbitrary space point onto the nearest surface point. Using the obtained triangles as the base, we then construct SES and SAS elements and derive the formulas for estimating the parameters of these surface elements, i.e., their coordinates, normals, and areas.

The described algorithm was first applied within the framework of the TAGSS (Triangulate Area Grid of Smooth Surface) software [5–7]. It differs from the above described algorithms by some specific scenarios of overcoming the conflicts arising upon the addition of new triangles. However, error in the scenarios of adding a new triangle led to frequent program failures. Moreover, there was no mechanism of the adaption of a grid to the presence of surface inflections and close, but non-adjacent surface domains. This also led to algorithm failures and excessively rough triangulation. These problems were overcome in the improved TAGSS version implemented as a subroutine of the DISOLV software [8–9, 35]. In these works, a method for constructing non-planar triangles on a torus and calculating their surface area was added, the case of several closed surfaces (several molecules or intracavity inclusions except inner cavities that was already taken into account in TAGSS) was considered, and the parameters of SAS elements were calculated.

In the given paper, we describe the algorithm in more details that have not been considered in the previous works [8–9, 35] and its further improvements. This algorithm was used as the base for developing the improved TAGSS program version implemented as a subroutine of the DISOLV software [8–9, 35].

## 2. Stages of the Triangulation of Molecular Surfaces

### 2.1. Input and Output Data of the Algorithm

The input data of the program are the arrays of atomic coordinates and van der Waals radii and the arrays of the parameters describing toroidal and spherical surface segments that appear after primary and secondary rolling. The maximum triangulation step is also specified.

The output data are the arrays of the structures of the data describing the triangulation grid on a molecular surface, the arrays of the parameters describing surface elements, and the surface area.

### 2.2. Determination of the Eigenbasis of a Molecule and Transformation into This Basis. Partitioning of the Outer and Inner Space of a Molecule into Overlapping Cubic Domains

The eigenbasis of a molecule is understood to mean the basis corresponding to a coordinate system, whose center is located at the geometric center of a molecule and axes are oriented along the principal axes of the geometric inertia tensor of a molecule.

The partitioning of the outer and inner space of a molecule into cubic domains represents the process of writing the indices of atoms, molecule surface segments, and surface elements into the set of the arrays, each of which corresponds to a cubic domain of specified size and certain position. We should note that cubic domains overlap each other.

The partitioning into cubic domains is performed to further simplify the procedures of searching for the index of an atom or surface element, which is nearest to a point with specified coordinates.

The details of this algorithm are described in [4].

## 2.3. Checking Whether All the Atoms Lie Within One and Only One Outer Molecular Surface

All the atoms must lie within one and only one outer molecular surface, which is found by the surface construction algorithm [4]. For this reason, it is advisable to check this property after finding molecular surfaces. To accomplish this, we can use the following integral:

$$\frac{1}{4\pi}\oiint \frac{\mathbf{r}-\mathbf{R}}{|\mathbf{r}-\mathbf{R}|^3}d\mathbf{S} = \begin{cases} 0, & \text{if an atom lies out of the integral surface} \\ 1, & \text{if an atom lies within the integral surface} \end{cases}, \quad (1)$$

where *R* is the radius vector of an atom.

## 2.4. Triangulation Algorithm
## 2.4.1. Projection Algorithm: Determining the Nearest Projection of a Specified Space Point onto the Surface of a Molecule

The estimated parameters of the principal surface fragments are used in the algorithm of determining the nearest projection of a specified space point onto the surface of a molecule, and this algorithm, in its turn, is applied to construct a surface triangulation grid.

For a specified space point, we determine the number of the cubic domain, within which it lies, and then begin to search for the nearest surface atoms within this cubic domain. If no any surface atoms have been found within this cubic domain, we begin the search among all the surface atoms. For the found nearest surface atom, we check whether the point enters one of its forbidden cones. If the point has not entered any forbidden cones, we construct the projection of this point onto the sphere of an atom. If such entrance takes place, we first perform the enumeration of the primary toroidal fragments surrounding the given atom.

For each enumerated toroidal primary fragment, we make an attempt to project the point onto its surface, fixing the distance between the initial and projection points. If the projection enters the domain forbidden by secondary rolling, we consider two convex secondary spherical segments of the two steady-state position spheres that bound a primary torus.

For each enumerated primary fragment (and secondary fragment, if required), we make an attempt to project the point onto its surface, fixing the distance between the initial and projection points. At the end of the enumeration of all the primary tori or contacting secondary fragments, we select the nearest projection from all the found projections.

If we have not managed to find such a projection, the enumeration of concave primary spherical elements surrounding the given atom is started. We make an attempt to project the point onto their surface, fixing the distance between the initial and projection points. If the projection enters the domain forbidden by secondary rolling, we first enumerate secondary toroidal elements and then convex secondary spherical elements surrounding the given primary segment.

For each enumerated primary fragment (and secondary fragment, if required), we make an attempt to project the point onto its surface, fixing the distance between the initial and projection points. At the end of the enumeration of all the primary spherical elements or contacting secondary fragments, we select the nearest projection from all the found projections.

We use the two types of projection: onto a toroidal fragment and onto a sphere with a specified center. If the center of a sphere $p_0$ with the radius $R$ and a certain point $r$ are specified, the projection onto the sphere $r_s$ is calculated as

$$r_s = p_0 + \frac{r - p_0}{|r - p_0|} R . \tag{2}$$

The normal of the surface $n_s$ at the point $r_s$ is determined by the expression

$$n_s = \pm \frac{p_0 - r_s}{|p_0 - r_s|}, \tag{3}$$

where the signs "+" and "−" are selected for concave and convex fragments, respectively.

For toroidal fragments, the projection of a point $r$ is calculated as

$$\mathbf{r}_s = \mathbf{p}_0 + \frac{\mathbf{r} - \mathbf{p}_0}{|\mathbf{r} - \mathbf{p}_0|} R_{pr},$$

$$\mathbf{p}_0 = \mathbf{p}_c + h(\cos(\alpha`)\mathbf{x} + \cos(\beta`)\mathbf{y}),$$

$$\cos(\alpha`) = \frac{((\mathbf{r} - \mathbf{p}_c) \cdot \mathbf{x})}{|\mathbf{r} - \mathbf{p}_c| \sin(\gamma`)}, \quad (4)$$

$$\cos(\beta`) = \frac{((\mathbf{r} - \mathbf{p}_c) \cdot \mathbf{y})}{|\mathbf{r} - \mathbf{p}_c| \sin(\gamma`)},$$

$$\cos(\gamma`) = \frac{((\mathbf{r} - \mathbf{p}_c) \cdot \mathbf{z})}{|\mathbf{r} - \mathbf{p}_c|}.$$

The angle $\alpha`$ is checked for whether its value lies within the interval ($\alpha\beta$) corresponding to the free rolling over a pair of atoms. For a primary rolling torus, we check whether the angle $\gamma`$ is smaller than the critical angle $\gamma$ determined by secondary steady-state position spheres (when secondary rolling is performed). For the normal, we use the equation

$$\mathbf{n}_s = \pm \frac{\mathbf{p}_0 - \mathbf{r}_s}{|\mathbf{p}_0 - \mathbf{r}_s|}, \quad (5)$$

where the signs "+" and "−" are used for primary and secondary tori, respectively.

### 2.4.2. Mechanisms of the Adaptive Adjustment of the Grid Step

Let $L$ be the adaptive grid step. The grid step is the radius $L$ of a circle circumscribed around an equilateral triangle with the height $Rch = 1.5L$. This triangle determines the maximum triangulation mesh. This maximum mesh $L^{max}$ is specified by a user. However, it must be changed near the areas, where two non-adjacent surface domains closely approach each other. Such areas are specified with the use of "adaption centers". An "adaption center" is the point, near which the change (adaption) of the standard maximum grid step is performed. The critical distance from "adaption centers" is the surface element–adaptation center distance, at which the grid step is changed. For each adaption center, we determine its maximum grid step. The adaption centers of a molecular surface of rolling [4] are classified into the following types:

(1) The centers of the narrowest areas (with a width δ) of primary tori (the centers of their "necks");

(2) The centers of the narrowest areas (with a width δ) of secondary tori (the centers of their "necks");

(3) The area of the closest approach (at a distance δ) of primary spherical segments; and

(4) The area of the closest approach (at a distance δ) of secondary spherical segments.

The critical distance from "adaption centers", i.e., the "adaption radius" is determined as $R^a \geq 2(Rch \cdot R_{mr})^{1/2}$, where $R_{mr}$ is the maximum curvature radius in the closest approach area. Note that $R_{mr} \leq R_{pr}$, where $R_{pr}$ is the primary rolling radius. For this reason, we may set $R^a = 2(Rch\ R_{pr})^{1/2}$.

The change of the grid step at a distance less than $R^a$ is given by the formula $L=\delta/4$.

Each cube is associated with all the "adaption centers" spaced from it at a distance that is less or equal to $R^a$.

It should be noted that the surface has fine structural features, which we do not want to loss due to a large grid step. In this connection, we reduce the grid step to sizes less than half principal curvature radii at the apices of the corresponding grid triangle. Moreover, all the apices of a triangle must lie either on the same segment or on the segments adjacent to each other (toroidal or spherical). The algorithm's section that provides these conditions will be described in more details in the triangulation scenarii.

### 2.4.3. Determining the Position of a Seed Triangle

The position of a seed triangle determines the starting point for the construction of a triangulation grid. From the viewpoint of our algorithm, this position may be selected arbitrarily. We specify a certain spatial point that is remote from a molecule and determine the corresponding projection of this point onto the surface. Then an equilateral triangle with a circumscribed circle radius that is nearly equal to the values specified by a user is constructed around the found projection point. The data on the first triangle are written into the arrays of the structures of the data describing a triangulation grid. The operation of the procedure of constructing a seed triangle results in the formation of a triangulation grid that consists of one triangle, three apices, and three boundary edges.

## 2.4.4. Triangulation Algorithm after the Creation of a Seed Triangle. Generation of a Triangulation Grid via the Sequential Addition of Triangles

### 2.4.4.1. Grid Generation Algorithm

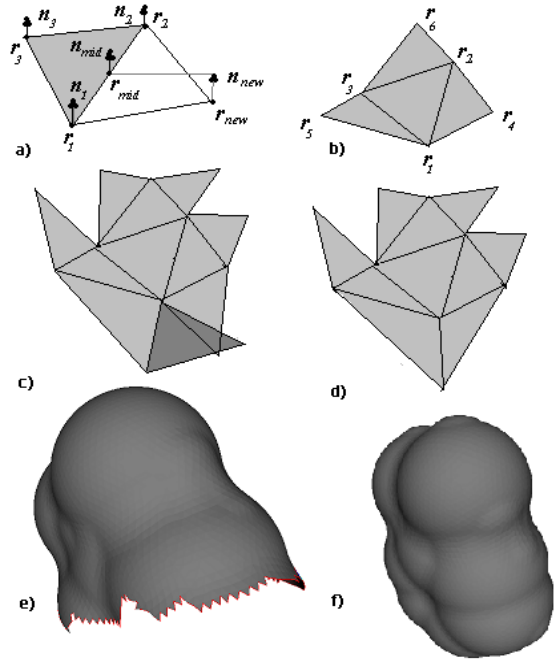

Fig. 1. Triangulation of a surface via the sequential addition of new triangles: (**a**) the addition of a new triangle and determination of a new surface point ($r_4$), which forms a new triangle together with the two end points ($r_1,r_2$) of the edge ($r_1$); (**b**) a surface fragment constituted of 4 triangles and a new set of boundary edges, which are obtained after detouring along all the edges of the first triangle; (**c**) the overlapping of triangles as a possible result of the sequential detour along the boundary edges and the addition of a new triangle to each boundary edge; (**d**) the joining of two triangles along a shared edge as one of scenarii applied to eliminate conflicts in the case of overlapping triangles; (**e**) the layer-by-layer construction of a surface via the repeated procedure of detouring along the boundary edges (each detour results in the formation of a new set of boundary edges, which may constitute several closed polygons); (**f**) the constructed surface (a surface is considered to be constructed, when no boundary edge remains).

The algorithm of generating a surface triangulation grid represents the cyclic process of passing along the boundary edges of a seed triangle with the addition of a new triangle to each boundary edge (Fig. 1). The operation of the algorithm is started from the passage along the seed triangle's boundary edges, which form the current set of boundary edges. In such a manner, a new set of boundary edges is formed, and this set forms the next set of current edges. When the processing of the current set is completed, the numbers of edges from the next set are written into the array of current boundary edges, thereupon all the indices in the array of the next boundary edges are set to zero. The cycle of processing is repeated again. During the addition of triangles, the cases of their overlapping and intersection are unavoidable. To resolve conflict situations arising in the process of adding a successive triangle, the geometric configuration of neighboring triangles is analyzed, and one of the possible methods of adding a triangle or modifying the geometric configuration of current boundary edges is selected. The current set of boundary edges may split into several sets of geometrically connected boundary edges, each of which forms a closed polygonal line. Hence, the boundary of already triangulated domains is a set consisting of one or several closed polygonal lines. The enumeration (cycle) is performed over the elements of this set and then over all the edges of polygons until this set of polygons disappears. At each step of these cycles, we apply one of the scenarii of adding a triangle or modifying the geometric configuration of current boundary edges. The grid generation process is stopped, when no boundary edges remain.

The operation of the procedure of generating a triangulation grid results in obtaining the filled arrays of the tree types of data structures (points (apices), edge (angles), and triangles) describing the triangulation grid and the connections between these elements.

### 2.4.4.2. Procedure of Detouring along the Edges of a Current Polygon. Cycle Enumeration of Current Edges and Selection of a Suitable Scenario (All the Possible Scenarios Are Listed Below the Description of the Algorithm)

(1) A current edge is selected as follows.

Let us assume that earlier, before a new triangle was added, a current polygon was splitted into two new polygons with a shared apex. Then one of the edges, whose one end is the shared point of these two formed polygons, is selected as a current edge. Such a selection is aimed at the spatial separation of these two polygons. Really, the addition of new triangles leads to the disappearance of their shared point. After the formation of two disconnected polygons, we stop the performed cycle of the processing of a current polygon and pass to the initial point of the enumeration of all polygons. If a polygon has not been splitted into two polygons, we take the edge that is next to the current edge from the current set of edges;

(2) Let the current set that incorporates the current edge consist of three edges. Then we use Scenario 10 and pass to step 1;

(3) Let the current set that incorporates the current edge consists of four edges. Let the quadrangle formed by these edges have a "special" apex with the two properties. First, only two edges originate from this apex. Second, this apex is supporting for only one triangle formed by these two edges. If the dihedral angle between this edge with the "special" apex and the edge with the opposite quadrangle apex is smaller than $\pi/6$, we use Scenario 9 and pass to step 1, and if this angle is larger than $\pi/6$, we use Scenario 8 and pass to step 1. If there is no "special" apex, we use Scenario 7 and pass to step 1;

(4) The angle between the current edge and one of its two adjacent edges from the current set is not small ($\alpha_1 > \pi/9$) and the angle between the current edge and the other adjacent edge is small ($\alpha_2 < \pi/9$). In this case, if the different ends of the current and adjacent edges that form a small angle are not connected by another path of two edges, we use Scenario 6 and pass to step 1. Otherwise, we use Scenario 5 and pass to step 1;

(5) If both angles $\alpha_1$ and $\alpha_2$ are small ($\alpha_1 < \pi/9$ and $\alpha_2 < \pi/9$), we use Scenario 1 and pass to step 1;

(6) We construct a "new" triangle and a "new" point by Scenario 2, but do not yet confirm their construction as a grid node and triangle;

(7) Let the angles $\beta_1$ and $\beta_2$ be the angles formed by the edges adjacent to the current edge and the corresponding "new" triangle's edges adjacent to them;

(8) Let at least one of the angles $\beta_1$ or $\alpha_1$ be small, and the angles $\beta_2$ and $\alpha_2$ be large ([($\beta_1<\pi/6$ or $\alpha_1<2\pi/9$) and ($\beta_2>\pi/6$ and $\alpha_2>2\pi/9$)]). Or, otherwise, at least one of the angles $\beta_2$ or $\alpha_2$ be small, and the angles $\beta_1$ and $\alpha_1$ be large [($\beta_2<\pi/6$ or $\alpha_2<2\pi/9$) and ($\beta_1>\pi/6$ and $\alpha_1>2\pi/9$)]). Then we apply Scenario 4 for the current and adjacent edges that form a small angle and return to step 1;

(9) Let the angles $\beta_1$ and $\beta_2$ be small ($\beta_1<\pi/6$ and $\beta_2<\pi/6$). Then we apply Scenario 4 for the current and adjacent edges that form a smaller angle and return to step 1.

(10) Let all the above defined angles $\beta_1$, $\alpha_1$, $\beta_2$, and $\alpha_2$ be large ($\beta_1>\pi/6$, $\alpha_1>2\pi/9$, $\beta_2>\pi/6$, and $\alpha_2>2\pi/9$). Let $\boldsymbol{n}_{mid}$ be the normal to the surface at the point of the current edge midpoint's projection, and $\boldsymbol{n}_{new}$ be the normal to the surface at the "new" point of a "new" triangle. Let $\gamma$ be the angle between $\boldsymbol{n}_{new}$ and $\boldsymbol{n}_{mid}$. If the angle $\gamma$ is large ($\gamma>\pi/2$), we apply Scenario 1 and return to step 1;

(11) Let at least one of the above defined angles $\beta_1$, $\alpha_1$, $\beta_2$, and $\alpha_2$ be small ($\beta_1<\pi/6$, or $\alpha_1<2\pi/9$, or $\beta_2<\pi/6$, or $\alpha_2<2\pi/9$). Then we perform Scenario 1 and return to step 1;

(12) The "first special" point that is close to the two apices of the current edge is found by the formula

$$\boldsymbol{r}^{(1)} = \frac{\boldsymbol{r}_1 + \boldsymbol{r}_2 + 2\boldsymbol{r}_{new}}{4}, \qquad (6)$$

where $\boldsymbol{r}_{new}$ is the radius vector of the "new" point, and $\boldsymbol{r}_1$ and $\boldsymbol{r}_2$ are the radii vectors of the two apices of the current edge. For all the points (in the cubic domain containing the "first special" point), we form the set of all the grid nodes that has the following properties:

(a) They lie in the cubic domain containing the "first special" point; and

(b) They do not coincide with the apices of the current edge and the apices of its two adjacent edges.

Further, we find the "second special" point. This points belongs to the found set and spaced from the "first special" point at the minimum distance $||r_{min}||$.

(13) If the distance $||r_{min}||>Rch$, we confirm the construction of a "new" triangle by Scenario 2 and return to step 1. If $\alpha_1<\pi/2$ or $\alpha_2<\pi/2$, we perform Scenario 1 and return to step 1;

(14) If the "second special" point does not lie on the current boundary, we perform Scenario 1 and return to step 1;

(15) If $\|r_{min}\|$ is smaller than half a current edge, we perform Scenario 1 and return to step 1;

(16) Let $n_m$ be the normal at the "second special" point. If the angle between $n_{new}$ and $n_m$ exceeds $\pi/2$, we perform Scenario 1 and return to step 1;

(17) We perform scenario 3. Two closed polygonal lines of the boundary edges with a shared point are thus formed. We return to step 1.

**2.4.4.3. Main Scenarii (Fig. 3) of Adding a New Triangle or Processing the Current Boundary Edges**

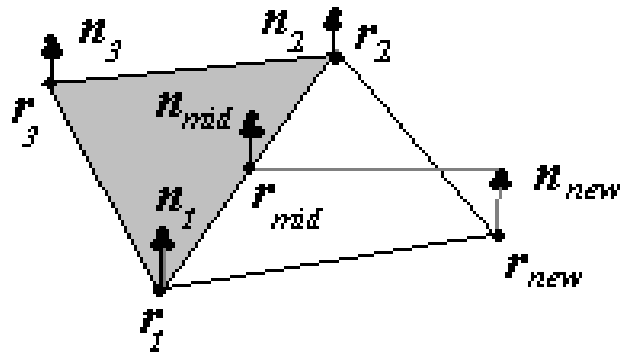

Fig. 2. Determining a new surface point ($r_{new}$), which forms a new triangle together with the edge's end points ($r_1, r_2$).

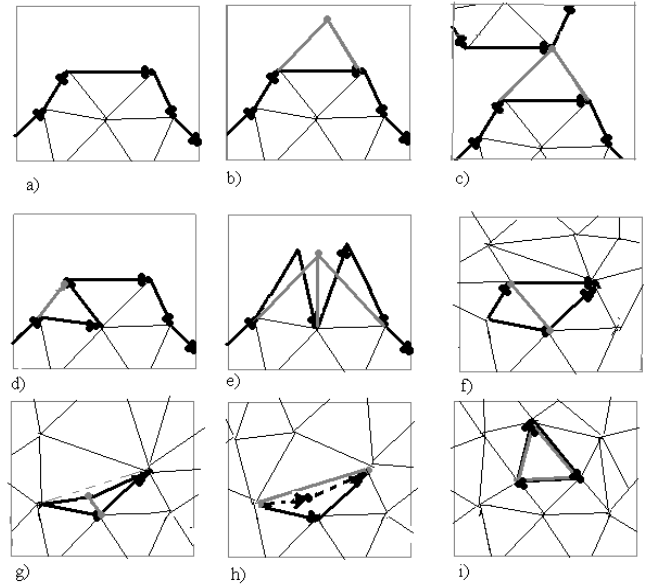

Fig. 3. Schematic illustration of the geometric configurations of a triangulation grid for the applied scenarii of the processing of boundary edges (solid arrows indicate the boundary edges of a current set and the direction of the detour along them, and dashed arrows indicate new boundary or inner edges): **(a)** the case when we does not do anything, if all other scenarios have failed to function (*Scenario 1*); **(b)** the formation of a triangle by a current boundary edge and two new boundary edges (dashed lines indicate new boundary edges) (*Scenario 2*); **(c)** the construction of a triangle, which geometrically connect or disconnect current boundary arrays (*Scenario 3*); **(d)** the formation of a triangle by the current and next (preceding) boundary edges of a current boundary set and a new boundary edge (dashed lines indicate a new boundary edge that forms a new triangle) (*Scenarii 4–5*); **(e)** the merging of two boundary edges that form an acute angle into one inner edge (*Scenario 6*); **(f)** the partitioning of a four-edge cycle into two triangle by the newly constructed fifth edge that connects the opposite obtuse angles of a quadrangle (*Scenario 7*); **(g)** the partitioning of a four-edge cycle into two triangles by the additional fifth edge that connects the "special" apex with the opposite apex (*Scenario 8*); **(h)** the elimination of the "special" apex and the formation of a triangle of the remaining three apices (*Scenario 9*); **(i)** the construction of a triangle that closes the boundary set of three edges (*Scenario 10*).

(1) We do not do anything, when all other scenarii have failed.

(2) We construct a "new" triangle supported by the current edge (Fig. 2) as follows. The midpoint of the current edge is projected onto the surface. At this point, we construct a vector so that it is perpendicular to the normal at this point and the vector of the current edge and have a length $Rch = 1.5L$. The obtained point is projected onto the surface. This is a "new" point. A "new" triangle is constructed from the current edge and the obtained "new" point.

$L$ is the adaptive grid step, which represents the radius of a circle circumscribed around an equilateral triangle with a height $Rch = 1.5L$. This triangle determines the maximum triangulation mesh in the given area.

The adaption of the grid step is performed by the following algorithm. If the cube determined by the midpoint of the current edge is not associated with any "adaption centers", this grid step is determined by the prespecified maximum size $L = L^{max}$. If such centers exist, the distance from each of these "adaption centers" associated with the cube to the midpoint of the current edge $||r_j||$ ($j=1,...,N_a$ is the number of an adaption center) is calculated. Among these distances, we find the distances that are less than the critical distance $||r_j|| < R^a_j$ for the corresponding "adaption center". For each similar "adaption center" $j$, there exist its own grid step $L = L^a_j$. Among them we select the minimum step $L = L^{min}$.

For each of the two edges of the current edge (though equally as for any surface point), there are the two principal curvature radii $R^{g1}$ and $R^{g2}$. For a torus, they are the radii of its two generatrices at this point (one of them has the same value everywhere and the other is increased from the center of a torus to its boundary). For a spherical segment, both principal radii are equal to its radius. The grid step is corrected so that it does not exceed half a radius for both apices as

$$L = min\left(L^{min}, R^{g1}_1/2, R^{g2}_1/2, R^{g1}_2/2, R^{g2}_2/2\right). \tag{7}$$

Then we construct a "new" point using such a grid step.

Let a "new" point lie on the segment $S_N$ (spherical or toroidal). The apices of the current edge lie on the segments $S_1$ and $S_2$. We check whether $S_N$ coincides with $S_1$ or $S_2$ or is neighboring for both these segments. For a "new" point, we find its two principal curvature radii $R^{g1}_N$ and $R^{g2}_N$ and check whether the conditions $L < R^{g1}_N/2$ and $L < R^{g2}_N/2$ are true. If at least one of these conditions is not true, we reduce the grid step as follows

$$L = min\left(L/2, R^{g1}_N/2, R^{g2}_N/2\right). \qquad (8)$$

Then we construct a "new" point using such a grid step and check the above described conditions again. This process is continued until these conditions are met.

(3) A "new" point constructed by method (2) (a) "merges" with the "second special" point (grid node) into one point, and (b) the "merging" occurs in compliance with the formula

$$r = \frac{(N_a r_a + N_b r_b)}{(N_a + N_b)}, \qquad (9)$$

where $N_a = 2$ is the number of edges originating from point (a), $N_b$ is the number of edges originating from point (b), $r_a$ and $r_b$ are the radii vectors of points (a) and (b), respectively, and $r$ is the radius vector of the point formed by merging;

(4) We construct a new triangle using, first, the current edge, second, the boundary edge that is adjacent to the current edge and forms a small angle with it and, third, the new boundary edge opposite to this small angle. Let point (a) be a "new" point constructed by Scenario 2. Second point (b) of the adjacent edge that does not belong to the current edge is shifted in compliance with the formula

$$r = \frac{(N_a r_a + N_b r_b)}{(N_a + N_b)}, \qquad (10)$$

where $N_a = 2$ and $N_b$ are the numbers of edges originating from point (a) and (b), respectively, $r_a$ is the radius vector of point (a), $r_b$ is the radius vector of the initial position of point (b), and $r$ is the radius vector of the new position of point (b);

(5) We construct a new triangle from the current edge, one of the boundary edges that are adjacent to the current edge and form a small angle with it, and the new boundary edge opposite to this small angle;

(6) The two adjacent boundary edges that have a shared point and form a small acute angle are merged into one inner (non-boundary) edge. Two different points of the adjacent edges (points (a) and (b)) are "merged" into one point. The radius vector of the resulting point is calculated as

$$r = (N_a r_a + N_b r_b) / (N_a + N_b), \tag{11}$$

where $N_a$ and $N_b$ are the numbers of edges originating from point (a) and (b), respectively, $r_a$ and $r_b$ are the radii vectors of point (a) and (b), respectively, and $r$ is the radius vector of the point formed by merging;

(7) The four-edge cycle is partitioned into two triangles by the newly constructed fifth edge connecting the opposite obtuse angles of a quadrangle;

(8) The four-edge cycle is partitioned into two triangles by the additional fifth edge connecting the "special" apex with its opposite apex;

(9) The "special" apex of the four-edge cycle is excluded. The remaining three apices are used to form a triangle;

(10) The triangle that closes the boundary set is constructed of three edges as follows. We perform a special check in order to prevent using this method at the first step, when the boundary set of a seed triangle is processed.

The first scenario is most important. Here we calculate the coordinates of the first apex. The method of this calculation is illustrated in Fig. 2.

The geometric configurations of the application of some important scenarios are schematically shown in Fig. 3. Boundary edges considered to be directed segments (oriented towards detouring along them).

## Final "Settling" of the Grid

To create a more uniform grid, we use the mechanism of "settling". We enumerate all the nodes of the grid. For all the nearest neighbors of a current node, we find the "centroid", which is then projected onto the surface, thus replacing this current node. The process of settling may be repeated several times.

## 2.4.5. Forming the Arrays of the Parameters of Surface Elements on the Basis of Surface Triangulation Data

Using the data on a surface triangulation grid as the base, we form the array of the data on surface elements. Each surface element is characterized by the spatial coordinates, surface area, and the direction of a normal to its surface. The coordinates and normal correspond to the nodes of the obtained triangulation grid. Each similar node is considered to be the center of a future polygonal surface element. The area of a triangular surface element is calculated by the following methods.

### 2.5.5.1. Formulas for the Areas of Triangular Surface Elements

Let us give the formulas for the areas of triangular surface elements. They may be classified into the three types: spherical, toroidal, and boundary elements, whose points belong to different surface fragments.

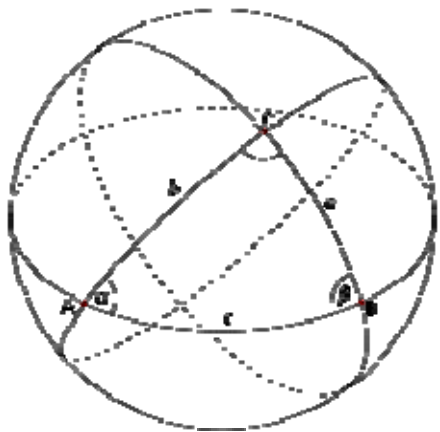

Fig. 4. Spherical triangle.

*(1) Surface Area of a Spherical Triangle (Fig. 4)*

The sum $s$ of the angles of a spherical triangle $\alpha, \beta$, and $\gamma$

$$s = \alpha + \beta + \gamma \tag{12}$$

is always lower than $3\pi$ and higher than $\pi$. The value

$$\varepsilon = s - \pi \tag{13}$$

is called the spherical excess. The surface area of a spherical triangle $S$ is determined by the Girard's formula

$$S = R^2 \varepsilon, \tag{14}$$

where R is the radius of a sphere.

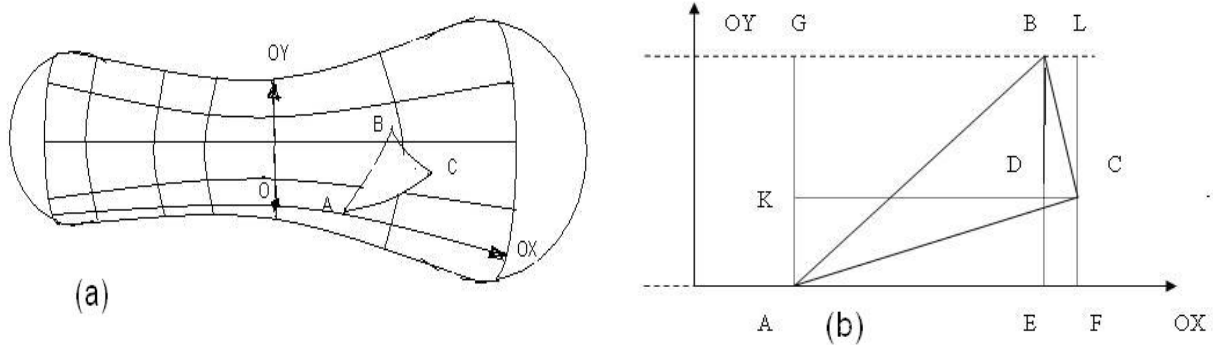

Fig. 5 (a) Toroidal triangle ABC and (b) its mapping onto a surface (proportions are distorted).

*(2) Surface Area of a Toroidal Triangle (Fig. 5)*

Let us consider a toroidal triangle ABC on a torus.

We shall determine a curvilinear coordinate system formed by the torus generatrices the two types: OX and OY are the axes directed along the generatrices (OY is represented by cricles perpendicular to the axis of a torus and OX is formed by the points of contact betweem the torus and a rolling sphere).

Let us take any three points A, B, and C. The OX and OY basis axes directed along the generatrices can always be selected as follows:

(1) Point C lies between points A and B along the OY axis;

(2) The positive direction of the OX axis is from segment [AB] to point C;

(3) For points A $(x_A, y_A)$ and B $(x_B, y_B)$, $x_B > x_A$ and $y_B > y_A$;

(4) The lines connecting the pairs of points AC, AB, and CB can not always be selected as coinciding with the geodesic lines of a torus. For this reason, let us select them as line segments between the points, which are the images of the geodesic lines of a cylinder and obtained via the mapping of a torus onto this cylinder. This mapping is performed as follows: the OX generatrices of a torus are mapped onto the cylinder's generatrices parallel to the cylinder's axes without changing their length, and the OY generatrices of a torus are mapped onto the cylinder's generatrices perpendicular to the cylinder's axes without changing their angular sizes;

(5) $\alpha_A$ is the angular position along the OX generatrix, and $x_A = R_{ox}\alpha_A$, where $R_{OX}$ is the torus rolling sphere radius, $h$ is the distance from the center of a rolling sphere to the straight line connecting the two supporting spheres of a torus;

(6) $\varphi_{AB}$ is the angular distance between points A and B along the OY generatrix;

Then the surface area of toroidal triangles and quadrangles is determied as follows.

The surface area of square EAGB is calculated as

$$S_{EAGB} = \varphi_{AB} R_{OX} (h\alpha - R_{OX} \sin(\alpha))|_{\alpha_A}^{\alpha_B} . \tag{15}$$

The surface areas of triangles ABE, BCD, and ACF are found as

$$S_{ABE} = \varphi_{AB} \frac{R_{OX}}{\alpha_B - \alpha_A} (h\frac{(\alpha - \alpha_A)^2}{2} - R_{OX}((\alpha - \alpha_A)\sin(\alpha) + \cos(\alpha)))|_{\alpha_A}^{\alpha_B} , \tag{16}$$

$$S_{BCD} = \varphi_{BC} \frac{R_{OX}}{\alpha_B - \alpha_C} (h\frac{(\alpha - \alpha_C)^2}{2} - R_{OX}((\alpha - \alpha_C)\sin(\alpha) + \cos(\alpha)))|_{\alpha_B}^{\alpha_C} , \tag{17}$$

$$S_{ACF} = \varphi_{AC} \frac{R_{OX}}{\alpha_A - \alpha_C} (h\frac{(\alpha - \alpha_C)^2}{2} - R_{OX}((\alpha - \alpha_C)\sin(\alpha) + \cos(\alpha)))|_{\alpha_A}^{\alpha_C} . \tag{18}$$

The surface area of triangle ABC is determined as

$$S_{ABC} = S_{ABE} - S_{ACF} + sign(x_C - x_B)(S_{EDCF} + S_{BCD}) . \tag{19}$$

*(3) Surface Area of the Triangles, Whose Apices Do Not Belong to the Same Surface Fragment*

Let a triangle be formed by the apices $r_1, r_2,$ and $r_3$. We shall use the formula for the surface area of a planar triangle to determine the surface area $S$ of such triangles.

The edges $a$ and $b$ of a triangle are determined as

$$\mathbf{a} = \mathbf{r}_2 - \mathbf{r}_1 . \text{and } \mathbf{b} = \mathbf{r}_3 - \mathbf{r}_1 \tag{20}$$

The surface area $S$ of a triangle is found as

$$S = \frac{1}{2}|[\mathbf{a} \times \mathbf{b}]| . \tag{21}$$

## 2.5.5.2. Formation of Polygonal Surface Elements and Calculation of Their Parameters

*Polygonal* surface elements are formed on the basis of triangular surface elements. The centers of polygonal elements coincide with the apices of triangular elements. The apices of polygonal elements are the midpoints of the edges, which belong to triangular elements and originate from the centers of polygonal elements, and the centroids (median intersection point) of these triangular elements.

The surface area of a polygonal element $S^M$ is calculated by summing one thirds of the surface areas $S^{tr}_i$ of triangular elements with the apix at the center of a polygonal element. For each triangulation grid node, we sum the surface areas of the triangles, for which the given node is an apix. The resulting total surface area is divided by 3, i.e.,

$$S^M = \frac{\sum_i S^{tr}_i}{3}, \qquad (22)$$

and the obtained result is memorized as the area of a surface element with the coordinates and normal that correpond to the given triangulation grid node.

## 2.5.5.3. SAS Elements

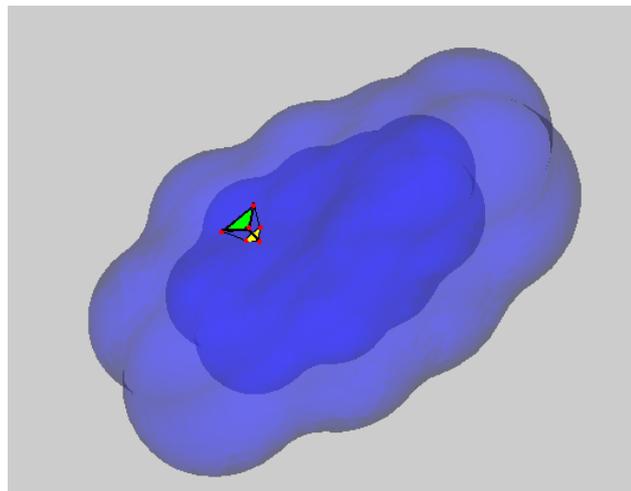

Fig. 6. Transformation of SES into SAS.

Let us have a surface after primary rolling and a triangle ($r_1, r_2, r_3$) on SES.
The image of this triangle on SAS ($r_{1n}, r_{2n}, r_{3n}$) (Fig. 6) is determined as

$$r_{1n} = r_1 + n_1 R_{pr}, \tag{23}$$

$$r_{2n} = r_2 + n_2 R_{pr}, \tag{24}$$

$$r_{3n} = r_3 + n_3 R_{pr}, \tag{25}$$

where $R_{pr}$ is the primary rolling radius, and $n_1$, $n_2$, and $n_3$ are the normals at corresponding points.

Then

(1) A surface element on a primary sphere supported by three atoms is mapped into a point. Its surface area on SAS is equal to zero;

(2) A surface element on a torus supported by two atoms is mapped into a line. Its surface area on SAS is also equal to zero;

(3) A surface element on an atom is mapped into a spherical triangle. Its surface area on SAS is proportional to its surface area on SES, i.e., $S_{ses}((R_{atom} + R_{pr})/R_{atom})^2$;

(4) A boundary surfce element between a torus and a pirmary sphere supported by three atoms has a zero surface area; and

(5) A boundary triangular surface element supported by a primary torus and an atom, a primary sphere and an atom, and by a primary torus and sphere and an atom has the surface area $S$ calculated by the formula for three points ($r_{1n}, r_{2n}, r_{3n}$) as

$$a = r_{2n} - r_{1n}, \tag{26}$$

$$b = r_{3n} - r_{1n}, \tag{27}$$

$$S = |a \times b|/2. \tag{28}$$

### 2.5.5.4. Calculating the Areas and Volume of SES and SAS

The area of a surface is calculated by summing the areas of all surface elements as

$$S = \oint_S dS. \tag{29}$$

The volume of a surface is determined as

$$V = \frac{1}{3}\oint_S (r \cdot n) dS, \tag{30}$$

where $r$ is the radius vector of the current point of a surface, and $n$ is its normal.

## 3. Conclusions

The represented algorithm and software developed on the basis if this algorithm allows us to quickly and reliably perform the triangulation of a smooth surface. The surface proves to be colored depending on the type of the nearest atoms, thus providing a convenience for visualization, and triangulated, thus allowing us not only to calculate its area and the volume enveloped by this surface, but also to solve the integral equation used in the continual model of a solvent with a rather high precision. The surface obtained via this program may be used both with the purpose of molecular visualization, which is especially topical for large albumin molecules, and for the calculation of the solvation contributions to the intermolecular interaction energy in the presence of a surrounding medium.

The triangulation algorithm can easily be generalized to the case of any smooth surface, including a smooth level surface.

## Acknowledgments

The given study is a more detailed presentation and further development of the work [35]. We are profoundly grateful to all the authors for their studies taken as the basis for the given paper.

# КОНТИНУАЛЬНАЯ МОДЕЛЬ СРЕДЫ II: УНИВЕРСАЛЬНЫЙ АДАПТИВНЫЙ АЛГОРИТМ ДЛЯ ТРИАНГУЛЯЦИИ ГЛАДКОЙ МОЛЕКУЛЯРНОЙ ПОВЕРХНОСТИ

Купервассер* О.Ю., Ваннер** Н.Э.

* ООО «Транзист Видео», участник Сколково

**Государственное научное учреждение Всероссийский научно-исследовательский институт ветеринарной санитарии, гигиены и экологии Россельхозакадемии, Москва

*E-mail: olegkup@yahoo.com

В данной работе представлен алгоритм, описывающий оригинальные и универсальные принципы триангуляции гладкой молекулярной поверхности, исключающей растворитель (SES), полученной первичной и вторичной обкаткой, и производной от нее поверхности, доступной растворителю (SAS). Эти поверхности играют роль границы между областями молекулы и растворителя.

Оригинальность данной работы состоит в создании универсального и адаптивного алгоритма триангуляции. Универсальность алгоритма триангуляции заключается в том, что он пригоден не только для поверхности обкатки, состоящей из фрагментов торов и сфер, но любой гладкой поверхности, в том числе и для любой поверхности уровня. Адаптивность этого алгоритма заключается в том, что размер элемента триангуляции может меняться в зависимости от его местоположения, отражая даже небольшие, но гладкие особенности поверхности; предотвращая «перескок» на несоседние, но близкие участки поверхности, исключая «обрезание» узких перешейков и каналов. Это достигается уменьшением шага сетки триангуляции до величины меньшей двух главных радиусов кривизны поверхности, а также вблизи «активного центра» - места сближения на малое расстояние несоседних участков поверхности.

Полученная триангулированная поверхность может быть использована для демонстрационных целей в молекулярных редакторах (сам алгоритм подходит для триангуляции любой гладкой поверхности, например поверхности уровня) а также и для расчета энергии сольватации для континуальных моделей растворителя.

Ключевые слова: триангуляция, адаптивность, молекулярная поверхность, первичная обкатка, вторичная обкатка

**1. Введение.**

В данной работе описывается универсальный адаптивный алгоритм триангуляции поверхности и формирования поверхностных элементов. Этот алгоритм будет использоваться в дальнейшем для расчета градиента энергии сольватации по сдвигам атомов на основе нахождения градиентов параметров поверхностных элементов. Важной особенностью и отличием данного алгоритма от многих других алгоритмов триангуляции молекулярной поверхности является то, что границы треугольных поверхностных элементов, лежащих вблизи границ тороидальных и сферических сегментов молекулярной поверхности, в общем случае не лежат на границе между этими сегментами. Это значительно упрощает расчет градиентов параметров таких граничных поверхностных элементов.

Приведем краткий обзор работ, посвященных триангуляции поверхности молекулы.
Для построения поверхности молекула субстрата все его атомы окружаются сферами радиуса Ван-дер-Ваальса [1-3]. Для получения гладкой поверхности, необходимой для сходимости метода, обычно производится первичная и вторичная «обкатка» поверхности сферами. Затем производиться триангуляция поверхности – разбиение на маленькие треугольники. Из этих треугольников в дальнейшем и формируются поверхностные элементы.

Существует два типа гладких поверхностей, окружающих молекулу, построение которых было подробно обсуждено в предыдущих работах [4-9]:

SAS (Solvent Accessible Surface) - поверхность доступная растворителю образуется центрами молекул растворителя, касающихся молекулы субстрата.

SES (Solvent Excluded Surface) - поверхность исключённого из растворителя объёма. Объем, занимаемый растворителем, лежит *вне* объема, ограниченного этой поверхностью. Сам субстрат полностью лежит *внутри* этого объема

Дальнейшими шагами является триангуляция поверхности и построение на основе этих треугольников поверхностных элементов. Существует широкий класс алгоритмов, предназначенных для триангуляции молекулярных поверхностей. Большая часть из них не является универсальной и привязана к конкретному алгоритму построения поверхности и типу образующихся сегментов поверхности [10-19]. Универсальные алгоритмы «Марширующие Кубы» (Marching Cubes (MC)) и его усовершенствованные аналог «Дуальное оконтуривание» (Dual contouring (DC)) не учитывают индивидуальных свойств гладкости поверхности, что осложняет построение равномерной сетки [20-22 ].

В данной работе приводится универсальный алгоритм триангуляции, пригодный не только для триангуляции поверхности обкатки, но и любой поверхности уровня. Он относится к классу алгоритмов – метод распространяющегося фронта (Advancing front technique (AFT)) с дискретизацией поверхности напрямую (Direct surface meshing) [23-28]. Иные используемые названия - метод распространения через предиктор-корректор (continuation method through the predictor-corrector) [29-32], или последовательный процесс формирования дискретных элементов (Mesh Generation using Processing Sequences) [33] или метод «мощения» (Paving: method) [34].

Генерирование триангуляционной сетки осуществляется методом последовательного добавления треугольников на поверхность. При этом используется алгоритм проецирования произвольной точки пространства на ближайшую точку поверхности. В дальнейшем на основе полученных треугольников строятся поверхностные элементы на SES и SAS и приводятся формулы, определяющие параметры этих поверхностных элементов – их координаты, нормали и площади.

Впервые описываемый алгоритм был применен в рамках программы TAGSS (Triangulate Area Grid of Smooth Surface) [5-7]. Он отличается от уже описанных выше алгоритмов некоторыми специфическими сценариями преодоления конфликтов, возникающих при добавлении новых треугольников. Однако ошибки в сценариях добавления нового треугольника приводили к частым сбоям программы. Кроме того, отсутствовал механизм адаптации сетки к изгибам поверхности и наличию близких, но несоседних участков поверхности. Это также приводило к сбоям алгоритма и излишней грубости триангуляции. Эти проблемы были преодолены в рамках усовершенствованной версии программы TAGSS, вошедшей как часть в программу DISOLV [8-9,35]. В этих работах был также добавлен метод построения расчета площади неплоских треугольников на торе, рассмотрен случай нескольких замкнутых поверхностей (кроме внутренних полостей, уже рассмотренных в TAGSS, это случаи нескольких молекул или включений в полости), проведен расчет параметров поверхностных элементов на SAS.

В данной работе мы описываем подробности алгоритма, не изложенные в предыдущих работах [8-9,35] и его дальнейшие усовершенствования. На основе данного алгоритма были создана усовершенствованная версия программы TAGSS, вошедшая как часть в программу DISOLV [8-9,35]

## 2. Этапы триангуляции поверхностей молекул.

### 2.1 Входные и выходные данные алгоритма.

Входными данными для программы являются массивы координат и Ван-дер-ваальсовых радиусов атомов, а также массивы параметров, описывающих поверхностные тороидальные и сферические сегменты, возникшие в результате первичной и вторичной обкатки. Задается также максимальный шаг триангуляции.

Выходными данными являются массивы структур данных описывающих сетку триангуляции поверхности молекулы, массивы параметров описывающих поверхностные элементы, значение площади поверхности.

### 2.2 Определение собственного базиса молекулы и переход в него. Разбиение пространства вокруг и внутри молекулы на перекрывающиеся кубические области.

Под собственным базисом молекулы понимается базис, соответствующий системе координат с центром в геометрическом центре молекулы и осями, направленными вдоль главных осей геометрического тензора инерции молекулы.

Разбиение пространства вокруг и внутри молекулы на кубические области представляет собой процесс запоминания индексов атомов, поверхностных сегментов молекулы и поверхностных элементов в наборе массивов, каждый из которых соответствует кубической области заданного размера и определённого положения. Следует отметить, что кубические области перекрываются между собой

Разбиение на кубические области производиться с целью упрощения в дальнейшем процедур поиска индекса атома или элемента поверхности ближайшего к точке с заданными координатами.

Подробности этого алгоритма изложены в работе [4]

## 2.3 Проверка, что Все атомы лежат внутри одной и только одной из внешних молекулярных поверхностей

Все атомы должны лежать внутри одной и только одной из внешних молекулярных поверхностей, которую находит алгоритм построения поверхности [4]. Поэтому после нахождения молекулярных поверхностей разумно совершить проверку этого свойства. Мы можем для этого использовать следующий интеграл:

$$\frac{1}{4\pi}\oint\frac{\boldsymbol{r}-\boldsymbol{R}}{|\boldsymbol{r}-\boldsymbol{R}|^3}d\boldsymbol{S} = \begin{cases} 0 & \text{если атом вне интегральной поверхности} \\ 1 & \text{если атом внутри интегральной поверхности} \end{cases}, \quad (1)$$

где $\boldsymbol{R}$ – радиус-вектор атома.

## 2.4 Алгоритм триангуляции.

### 2.4.1 Алгоритм проецирования: для заданной точке в пространстве определение ближайшей проекции на поверхность молекулы.

Определённые параметры основных фрагментов поверхности применяются в алгоритме определения ближайшей проекции заданной точки в пространстве на поверхность молекулы, который, в свою очередь, используется для построения сетки триангуляции поверхности.

Для заданной точки в пространстве определяется номер кубической области, в которой она лежит, далее производиться поиск ближайших поверхностных атомов из этой кубической области. Если поверхностных атомов в кубической области найдено не было – поиск среди всех поверхностных атомов. Для найденного ближайшего поверхностного атома производиться проверка попадания точки в один из его запрещённых конусов. Если точка не попала ни в один из запрещённых конусов ближайшего атома, то производиться проекция этой точки на сферу атома. Если же такое попадание имеет место, то производится перебор сначала первичных тороидальных окружающих данный атом.

Для каждого их перебираемых тороидальных первичных фрагментов производиться попытка проецировать точку на его поверхность, с фиксированием расстояния между исходной точкой и точкой проекции. Если проекция попадает в область, запрещенной вторичной обкаткой, то рассматриваются два выпуклых вторичных сферических сегмента двух сфер устойчивого положения, ограничивающих первичный тор.

Для каждого их перебираемых первичных фрагментов (а в случае необходимости и вторичных фрагментов) производиться попытка проецировать точку на его поверхность, с фиксированием расстояния между исходной точкой и точкой проекции. В конце перебора всех первичных торов или соприкасающихся вторичных фрагментов, выбирается наиболее близкая проекция из всех найденных.

Если таковую найти не удалось, начинается перебор вогнутых первичных сферических элементов, окружающих данный атом. Производиться попытка проецировать точку на их поверхность, с фиксированием расстояния между исходной точкой и точкой проекции. Если проекция попадает в область, запрещенной вторичной обкаткой, то производится перебор сначала вторичных тороидальных, а затем и выпуклых вторичных сферических элементов, окружающих данный первичный сегмент.

Для каждого их перебираемых первичных фрагментов (а в случае необходимости и вторичных фрагментов) производиться попытка проецировать точку на его поверхность, с фиксированием расстояния между исходной точкой и точкой проекции. В конце перебора всех первичных сферических элементов или соприкасающихся вторичных фрагментов, выбирается наиболее близкая проекция из всех найденных.

Используется два типа проецирования: проецирование на сферу с заданным центром и проецирование на тороидальный фрагмент. Если задан центр сферы $p_0$ радиуса $R$ и некоторая точка $r$, то проекция на сферу $r_s$ вычисляется по формуле:

$$r_s = p_0 + \frac{r - p_0}{|r - p_0|} R.$$

( 2 )

При этом нормаль поверхности $n_s$ в точке $r_s$ задаётся выражением:

$$n_s = \pm \frac{p_0 - r_s}{|p_0 - r_s|},$$

( 3 )

где знак «+» выбирается для вогнутых фрагментов, а знак «-» для выпуклых. Для тороидальных фрагментов определение проекции точки $r$ вычисляется по формулам:

$$r_s = p_0 + \frac{r - p_0}{|r - p_0|} R_{pr},$$
$$p_0 = p_c + h(cos(\alpha`)x + cos(\beta`)y),$$
$$cos(\alpha`) = \frac{((r - p_c) \cdot x)}{|r - p_c| sin(\gamma`)},$$
$$cos(\beta`) = \frac{((r - p_c) \cdot y)}{|r - p_c| sin(\gamma`)},$$
$$cos(\gamma`) = \frac{((r - p_c) \cdot z)}{|r - p_c|}.$$

( 4 )

При этом производиться проверка значения угла $\alpha`$ - лежит – ли оно внутри интервала ($\alpha\beta$) соответствующего свободной обкатке вокруг пары атомов. Для первичного тора обкатки проверяется (при наличии вторичной обкатки), что угол $\gamma`$ меньше критического угла $\gamma$, определяемого вторичными сферами устойчивого положения. Для нормали используется выражение:

$$n_s = \pm \frac{p_0 - r_s}{|p_0 - r_s|},$$

( 5 )

где + для первичных торов, а − для вторичных торов.

### 2.4.2 Механизмы адаптивного регулирования шага сетки.

Пусть $L$ – адаптивный шаг сетки. Шаг сетки - это радиус окружности $L$, описанной вокруг равностороннего треугольника с высотой, равной $Rch = 1.5L$. Этот треугольник определяет максимальный размер триангуляции. Этот максимальный размер $L^{max}$ задается пользователем. Однако он должен меняться вблизи тех мест, где два несоседних участка поверхности подходят близко друг к другу. Такие места задаются с помощью «центров адаптации». «Центр адаптации» – это точка, вблизи которой происходит изменение (адаптация) стандартного максимального шага сетки. Критическое расстояние от «центров адаптации» – расстояние от поверхностного элемента до центра адаптации, при котором происходит изменение шага сетки. Для каждого центра адаптации определяется его максимальный шаг сетки. Центры адаптации для молекулярной поверхности обкатки [4] подразделяются на следующие виды

1) центры самых узких мест (шириной δ) первичных торов (центр их «перешейков») .

2) центры самых узких мест (шириной δ) вторичных торов (центр их «перешейков») .

3) место максимального сближения (на расстояние δ) первичных сферических сегментов.

4) место максимального сближения (на расстояние δ) вторичных сферических сегментов.

Критическое расстояние от «центров адаптации» - т.е. «радиус адаптации» определяется следующей формулой:

$R^a \geq 2(Rch \cdot R_{mr})^{1/2}$, где $R_{mr}$ – максимальный радиус кривизны в месте максимального сближения. Заметим, что $R_{mr} \leq R_{pr}$, где $R_{pr}$ первичный радиус обкатки. Поэтому можно положить $R^a = 2(Rch \cdot R_{pr})^{1/2}$.

Изменение шага сетки на расстоянии меньшего $R^a$ дается формулой: $L = \delta/4$

Каждому кубу соотносятся все «центры адаптации» лежащие от него на расстоянии меньшем или равным $R^a$.

Следует отметить, что поверхность имеет тонкости структуры, которые мы не хотим потерять из-за большого шага сетки. В связи с этим, мы уменьшаем шаг сетки до размеров меньших половины главных радиусов кривизны в вершинах соответствующего треугольника сетки. Кроме того, все вершины одного треугольника должны лежать либо на одном и том же, либо на соседних по отношению друг к другу сегментах (тороидальных или сферических). Подробнее часть алгоритма, обеспечивающая эти условия, будет описана в сценариях триангуляции.

### 2.4.3 Определение положения затравочного треугольника.

Положение затравочного треугольника определяет начало построения сетки триангуляции. С точки зрения алгоритма построения это положение может быть выбрано произвольно. Задаётся некоторая удалённая от молекулы точка в пространстве, для которой определяется соответствующая проекция на поверхность. Затем вокруг найденной точки проекции выстраивается равносторонний треугольник с размером радиуса описанной окружности примерно равной величине заданной пользователем. Данные о первом треугольнике помещаются в массивы структур данных описывающих сетку триангуляции. В результате работы процедуры построения затравочного треугольника образуется сетка триангуляции, состоящая из одного треугольника, трёх вершин и трёх граничных рёбер.

### 2.4.4 Алгоритм триангуляции после создания затравочного треугольника. Генерирование сетки триангуляции методом последовательного добавления треугольников.

#### 2.4.4.1 Алгоритм генерирования сетки имеет следующий вид:

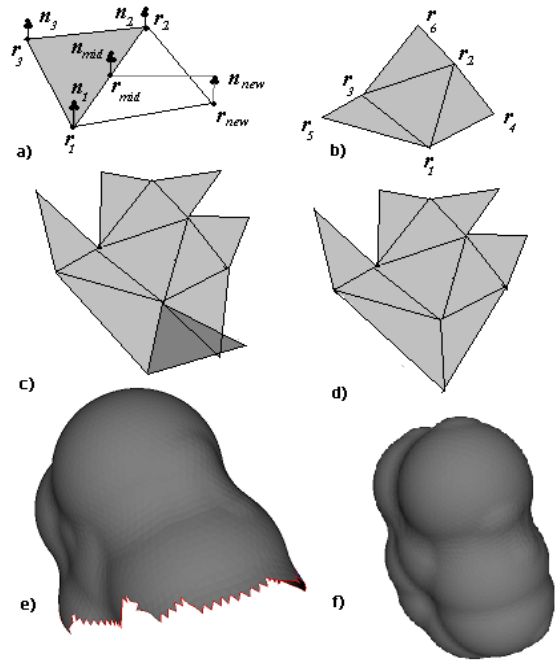

**Рис.1** Описание метода триангуляция поверхности последовательным добавлением новых треугольников.
**(a)** Добавление нового треугольника: Определение новой точки поверхности ($r_4$),

которая вместе двумя концевыми точками ($r_1$,$r_2$) ребра ($r_1$) образует новый треугольник.

**(b)** После обхода всех рёбер первого треугольника получаем фрагмент поверхности из 4

треугольников и новое множество граничных рёбер.

**(c)** Последовательно производя обход граничных рёбер и добавляя к каждому по новому треугольнику может возникнуть наложение треугольников.

**(d)** Для случаев наложения треугольников применяется ряд сценариев для устранения конфликтов,

например, сшивание двух треугольников общим ребром.

**(e)** Повторяя процедуру обхода граничных ребер, выполняется послойное построение поверхности. После каждой процедуры обхода образуется новый набор граничных рёбер. Граничные рёбра могут образовывать несколько замкнутых ломаных.

**(f)** Построенная поверхность. Поверхность считается построенной, когда не остаётся ни одного граничного ребра.

Алгоритм генерирования сетки триангуляции поверхности (Рис. 1) представляет собой циклический процесс обхода граничных рёбер с добавлением к каждому граничному ребру нового треугольника. Работа алгоритма начинается с обхода граничных рёбер затравочного треугольника, которые образуют текущий набор граничных рёбер. При этом формируется новый набор граничных рёбер, который формирует последующий набор текущих рёбер. По завершении обработки текущего набора, номера рёбер из последующего набора перемещаются в массив текущих граничных рёбер, после чего массив индексов последующих граничный рёбер обнуляется. Цикл обработки повторяется заново. В процессе добавления треугольников неизбежно возникают случаи их наложения и пересечения. Для разрешения конфликтных ситуаций в процессе добавления очередного треугольника производится анализ геометрической конфигурации соседних треугольников и выбор одного из возможных способов добавления треугольника или видоизменения геометрической конфигурации текущих граничных рёбер. При этом текущее множество граничных ребер может разбиваться на несколько множеств геометрически связанных граничных рёбер, каждое из которых образует замкнутую ломанную. Таким образом, граница уже триангулированных областей – множество, состоящее из одной или нескольких замкнутых ломаных. Делаем перебор (цикл), по элементам этого множества ломаных, затем перебор (цикл) по всем ребрам ломаных до тех пор, пока это множество ломаных не исчезнет. На каждом шаге этих циклов мы применяем один из сценариев добавления треугольника или видоизменения геометрической конфигурации текущих граничных рёбер. Процесс генерирования сетки оканчивается тогда, когда не остаётся ни одного граничного ребра.

В результате работы процедуры для генерирования сетки триангуляции получаем заполненные массивы трёх типов структур данных (Point (вершины), Edge (углы), Triangle (треугольники)), описывающих сетку триангуляции и связи между этими элементами.

### 2.4.4.2 Процедура обхода рёбер текущей замкнутой ломаной. Цикл перебора текущих ребер и выбор подходящего сценария (список всех возможных сценариев приведен ниже описания алгоритма):

1. Выбираем текущее ребро следующим образом:
Предположим что ранее, после добавления нового треугольника текущая замкнутая ломаная разделилась на две новые замкнутые ломаны с одной общей вершиной. Тогда в качестве текущего ребра выбираем одно из ребер, имеющих одним из концов общую точку этих двух образовавшихся ломаных. Целью такого выбора является пространственное разделение двух ломаных. Действительно, добавление новых треугольников приводит к исчезновению их общей точки. После образования двух не связанных ломаных прекращаем выполняемый цикл обработки текущей ломаной и идем на начальный пункт перебора всех ломаных. Если разделения ломаной на две не произошло, то берем следующее ребро после текущего в текущем множестве ребер.

2. Пусть текущее множество, включающее текущее ребро, состоит из трех ребер. Тогда используем сценарий 10 (и идем на пункт 1).
3. Пусть текущее множество, включающее текущее ребро, состоит из четырех ребер.

Пусть у образованного ими четырехугольника имеется «особая» вершина, обладающая двумя свойствами. Во-первых, из нее выходит только два ребра. Во-вторых, на нее опирается только один треугольник, образованный этими двумя ребрами. Если двухгранный угол между этой треугольной гранью с «особой» вершиной и гранью с противоположной ей вершиной четырехугольника меньше $\pi/6$, то используем сценарий 9 (и идем на пункт 1), а если больше – то сценарий 8 (и идем на пункт 1). Если «особой» точки нет, то используем сценарий 7 (и идем на пункт 1).

4. Один из углов текущего ребра с одним из двух смежных ребер из текущего множества не маленький ($α_1 > \pi/9$), а с другим маленький ($α_2 < \pi/9$):

Если разные концы текущего ребра и смежного ребра, образующие малый угол не связанны еще одним путем из двух ребер, то используем сценарий 6 (и идем на пункт 1). Иначе используем сценарий 5 (и идем на пункт 1).

5. Оба угла $α_1$ и $α_2$ – маленькие ($α_1 < \pi/9$ и $α_2 < \pi/9$), то используем сценарий 1 (и идем на пункт 1).
6. Строим «новый» треугольник и «новую» точку по сценарию 2, но пока не подтверждаем их построение в качестве узла и треугольника сетки.

7. Пусть углы $\beta_1$ и $\beta_2$ - углы между рёбрами, смежными к текущему ребру, и соответствующими смежными к ним сторонами «нового» треугольника

8. Пусть хотя бы один из углов $\beta_1$ или $\alpha_1$ мал, а углы $\beta_2$ и $\alpha_2$ -велики. ($[(\beta_1<\pi/6$ или $\alpha_1<2\pi/9)$ и $(\beta_2>\pi/6$ и $\alpha_2>2\pi/9)])$. Или, наоборот, хотя бы один из углов $\beta_2$ или $\alpha_2$ мал, а углы $\beta_1$ и $\alpha_1$ -велики $[(\beta_2<\pi/6$ или $\alpha_2<2\pi/9)$ и $(\beta_1>\pi/6$ и $\alpha_1>2\pi/9)])$. Тогда применяем сценарий 4 для текущего ребра и смежного ребра, образующего малый угол (и идём на пункт 1).

9. Пусть углы $\beta_1$ и $\beta_2$ малы ($\beta_1<\pi/6$ и $\beta_2<\pi/6$). Тогда применяем сценарий 4 для текущего ребра и смежного ребра, образующего меньший угол (и идём на пункт 1).

10.. Пусть все определённые выше углы $\beta_1$, $\alpha_1$, $\beta_2$ и $\alpha_2$ велики ($\beta_1>\pi/6$, $\alpha_1>2\pi/9$, $\beta_2>\pi/6$ и $\alpha_2>2\pi/9$). Пусть $\boldsymbol{n}_{mid}$ - нормали к поверхности в точке проекции середины текущего ребра, $\boldsymbol{n}_{new}$ – нормаль к поверхности в «новой» точке «нового» треугольника. Пусть $\gamma$- угол между $\boldsymbol{n}_{new}$ и $\boldsymbol{n}_{mid}$. Если угол $\gamma$ велик ($\gamma>\pi/2$), то применяем сценарий 1 (и идём на пункт 1).

11. Пусть хотя бы один из определённых выше углов $\beta_1$, $\alpha_1$, $\beta_2$ и $\alpha_2$ мал ($\beta_1<\pi/6$, или $\alpha_1<2\pi/9$, или $\beta_2<\pi/6$, или $\alpha_2<2\pi/9$). Тогда выполняем сценарий 1 (и идём на пункт 1).

12. Находим «первую особую» точку, близкую к двум вершинам текущего ребра по формуле:

$$\boldsymbol{r}^{(1)} = \frac{\boldsymbol{r}_1 + \boldsymbol{r}_2 + 2\boldsymbol{r}_{new}}{4}, \tag{6}$$

$\boldsymbol{r}_{new}$ – радиус-вектор «новой» точки.

$\boldsymbol{r}_1$, $\boldsymbol{r}_2$ - радиус-вектора двух вершин текущего ребра. Если для всех точек (в кубической области, содержащей «первую особую» точку)

Формируем множество всех узлов сетки, которые обладают следующими свойствами:

   a. Лежат в кубической области, содержащей «первую особую» точку

   в. Не совпадают с вершинами текущего ребра и вершинами двух смежных с ним граничных рёбер

Находим «вторую особую» точку. Эта точка из найденного множества, имеющая минимальное расстояние $||r_{min}||$ до «первой особой» точки.

13. Если расстояние $||r_{min}||>Rch$, то подтверждаем построение «нового» треугольника по сценарию 2 (и идем на пункт 1). Если $\alpha_1<\pi/2$ или $\alpha_2<\pi/2$, то выполняем сценарий 1 (и идем на пункт 1).

14. Если «вторая особая» точка не лежит на текущей границе, то выполняем сценарий 1 (и идем на пункт 1).

15. Если $||r_{min}||$ меньше половины текущего ребра, то выполняем сценарий 1 (и идем на пункт 1).

16. Пусть $\boldsymbol{n}_m$ – нормаль во «второй особой» точке. Если угол между $\boldsymbol{n}_{new}$ и $\boldsymbol{n}_m$ больше $\pi/2$, то выполняем сценарий 1 (и идем на пункт 1).

17. Выполняем сценарий 3. Образуются две замкнутые ломаные граничных рёбер с общей точкой. Идем на пункт 1.

### 2.4.4.3 Основные сценарии (Рис. 3) добавления нового треугольника или обработки текущих граничных рёбер:

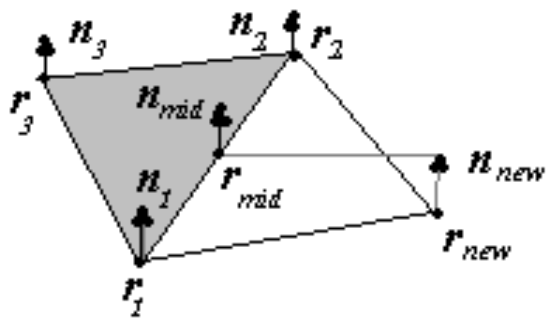

**Рис. 2.** Определение новой точки поверхности ($r_{new}$), которая вместе двумя концевыми точками ($r_1, r_2$) ребра образует новый треугольник.

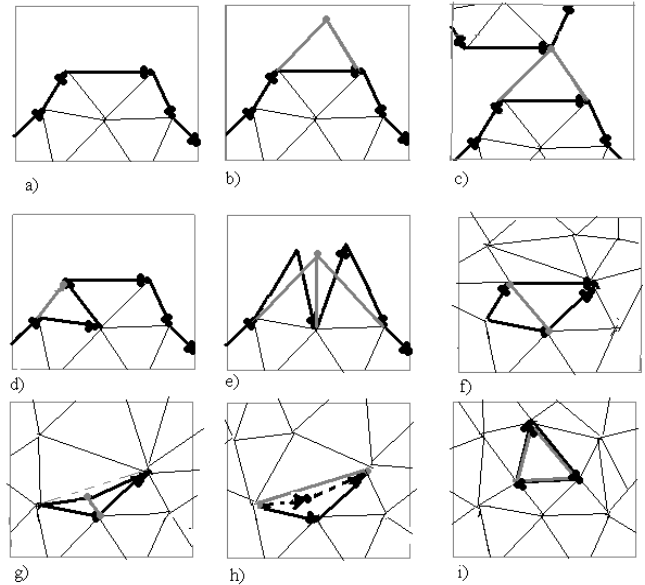

**Рис. 3**. Схематическая иллюстрация геометрических конфигураций сетки триангуляции для применяемых сценариев обработки граничных рёбер. Жирными отрезками со стрелками указаны граничные рёбра текущего набора и направление их обход. Пунктиром указаны новые граничные или внутренние рёбра.

**(a)** Ничего не делать. Когда все другие сценарии не сработали. *Сценарий 1*

**(b)** Треугольник образуется текущим граничным ребром и двумя новыми граничными рёбрами. Пунктиром показаны новые граничные рёбра. *Сценарий 2*

**(c)** Построение треугольника, геометрически объединяющего или разделяющего текущие граничные массивы. *Сценарий 3*

**(d)** Треугольник образуется текущим и последующим (предыдущим) граничным рёбрами текущего граничного набора и одним новым граничным ребром. Пунктиром показано новое граничное ребро образующее новый треугольник. *Сценарии 4-5*.

(**e**) Слияние двух граничных рёбер образующих острый угол между собой в одно внутреннее ребро. *Сценарий 6*.

(**f**) Цикл из четырех ребер разбивается на два треугольника новым построенным пятым ребром, соединяющий противоположные тупые углы четырехугольника. *Сценарий 7*

(**g**) Цикл из четырех ребер разбивается на два треугольника дополнительным пятым ребром, соединяющий «особую» вершину с противоположной ей вершиной. *Сценарий 8*

(**h**) «Особая» вершина цикла из четырех ребер выбрасывается. Из оставшихся трех вершин формируем треугольник. *Сценарий 9*

(**i**) Построение треугольника закрывающего граничное множество, состоящее из трёх рёбер. *Сценарий 10*

1) Ничего не делать. Когда все другие сценарии не сработали.

2) Строится «новый» треугольник, опирающийся на текущее ребро. (Рис. 2) Делается это следующим образом. Проецируем центр текущего ребра на поверхность. Из этой точки строим вектор, перпендикулярный нормали в этой точке и вектору текущего ребра и длинной равный *Rch =1.5L*. Проецируем полученную точку на поверхность. Это «новая» точка. Строим «новый» треугольник из текущего ребра и полученной «новой» точки.

*L* – это адаптивный шаг сетки. Адаптивный шаг сетки - это радиус окружности, описанной вокруг равностороннего треугольника с высотой *Rch =1.5L*. Этот треугольник определяет максимальный размер триангуляции в данном месте.

Адаптация шага сетки идет по следующему алгоритму. Если к кубу, определяемого серединой текущего ребра, не относится ни один «центр адаптации», то этот шаг сетки определяется заранее заданным максимальным размером $L =L^{max}$. Если такие центры есть, то считается расстояние от каждого из этих «центров адаптации», относящихся к кубу, до середины текущего ребра $||r_j||$, ($j=1,...,N_a$ – номер центра адаптации). Находим среди этих расстояний те, которые меньше критического для соответствующего «центра адаптации» $||r_j|| < R^a_j$. Каждому такому «центру адаптации» $j$ соответствует свой шаг сетки $L=L^a_j$. Выбираем среди них минимальный шаг $L=L^{min}$.

Каждой двух вершин текущего ребра (впрочем, как и любой точке поверхности) соответствует два главных радиуса кривизны $R^{g1}$, $R^{g2}$. Для тора это радиусы двух образующих тора в этой точке (один из них всюду одинаков, другой увеличивается от центра к краям тора). Для сферического сегмента оба главных радиуса равны его радиусу. Корректируем шаг сетки, чтобы он был не больше половины этих радиусов для обеих вершин:

$$L = min\left(L^{min}, R^{g1}_1/2, R^{g2}_1/2, R^{g1}_2/2, R^{g2}_2/2\right).$$

( 7 )

Далее строим «новую» точку с таким шагом сетки.

Пусть «новая» точка лежит на сегменте (тороидальном или сферическом) $S_N$. Вершины текущего ребра лежат на сегментах $S_1$, $S_2$. Проверяем, что $S_N$ либо совпадает с $S_1$ или $S_2$, либо является соседним для обоих из них. Для «новой» точки находим ее два главных радиуса кривизны $R^{g1}_N$, $R^{g2}_N$. Проверяем, что $L < R^{g1}_N/2$, $L < R^{g2}_N/2$. Если хотя бы одно из этих условий не выполняется, то шаг сетки уменьшается следующим образом:

$$L = min\left(L/2, R^{g1}_N/2, R^{g2}_N/2\right).$$

( 8 )

Далее строим «новую» точку с таким шагом сетки и снова проверяем описанные выше условия. Этот процесс продолжается до тех пор, пока эти условия не выполнятся.

3) «Новая» точка»-а, построенная по методу 2 «сливается» в одну точку со «второй особой» точкой (узлом сетки) - b. «Слияние» идет по формуле:

$$r = (N_a r_a + N_b r_b) / (N_a + N_b),$$

( 9 )

$N_a$ =2 - число ребер, выходящих из точки a; $N_b$ - число ребер, выходящих из точки b;

$r_a$ – радиус-вектор точки a; $r_b$ - радиус-вектор точки b; $r$ - радиус-вектор образующейся в результате слияния точки;

4) Строится новый треугольник. Он образуется, во-первых, текущим ребром. Во-вторых, граничным ребром, смежным к текущему ребру и имеющим с ним малый угол. И, в-третьих, одним новым граничным ребром, построенным напротив этого маленького угла. Пусть a - «новая точка», построенная по сценарию 2. Вторая точка b смежного ребра (не принадлежащая текущему ребру) сдвигается по формуле:

$$r = (N_a r_a + N_b r_b) / (N_a + N_b),$$

( 10 )

$N_a$=2 - число ребер, выходящих из точки a; $N_b$ - число ребер, выходящих из точки b;

$r_a$ – радиус-вектор точки a; $r_b$ - радиус-вектор начального положения точки b; $r$ - радиус-вектор нового положения точки b;

5) Строится новый треугольник, образуемый текущим ребром, одним из граничных рёбер, смежных к нему и имеющий с ним малый угол, и одним новым граничным ребром, построенным напротив этого маленького угла.

6) Слияние двух смежных граничных рёбер, имеющих общую точку и образующих малый острый угол между собой в одно внутреннее (не

граничное) ребро. Две различающиеся точки смежных ребер (точки a и b) «сливаются» в одну точку. Радиус-вектор получающейся точки считается по следующей формуле:

$$r = (N_a r_a + N_b r_b) / (N_a + N_b),$$

( 11 )

$N_a$ - число ребер, выходящих из точки a; $N_b$ - число ребер, выходящих из точки b;

$r_a$ – радиус-вектор точки a; $r_b$ - радиус-вектор точки b; $r$ - радиус-вектор образующейся в результате слияния точки;

7) Цикл из четырех ребер разбивается на два треугольника новым построенным пятым ребром, соединяющий противоположные тупые углы четырехугольника.

8) Цикл из четырех ребер разбивается на два треугольника дополнительным пятым ребром, соединяющий «особую» вершину с противоположной ей вершиной.

9) «Особая» вершина цикла из четырех ребер выбрасывается. Из оставшихся трех вершин формируем треугольник.

10) Построение треугольника закрывающего граничное множество, состоящее из трёх рёбер. Проводиться специальная проверка, чтобы не применять этот метод для первого шага – когда обрабатывается граничное множество затравочного треугольника.

Наиболее важен первый сценарий. Здесь производиться вычисление координат новой вершины. Метод этого вычисления пояснён на Рис. 2.

На Рис. 3 схематично представлены геометрические конфигурации применения некоторых важных сценариев. Граничные рёбра считаются направленными отрезками – в направлении, совпадающем с направлением их обхода.

**Окончательная «утряска» сетки.**

Для создания более однородной сетки используется механизм «утряски». Перебираются все узлы сетки. Для всех ближайших соседей текущего узла находится «центр тяжести». Затем он проецируется на поверхность и заменяет собой этот текущий узел. Процесс утряски может проводиться несколько раз.

### 2.4.5 Формирование массивов параметров поверхностных элементов на основании данных о триангуляции поверхности.

На основании данных о сетке триангуляции поверхности производиться формирование массивов данных о поверхностных элементах. Каждый поверхностный элемент характеризуется координатами в пространстве, направлением нормали поверхности и его площадью. Координаты и нормаль соответствуют узлам, полученной сетки триангуляции. Каждый такой узел рассматривается как центр будущего многоугольного поверхностного элемента. А для вычисления площади треугольного поверхностного элемента применяется следующие методы:

#### 2.5.5.1 Формулы для площадей треугольных поверхностных поверхностных элементов

Приведем формулы для площадей треугольных поверхностных элементов. Их можно разделить на три типа – сферические, тороидальные и граничные, чьи точки принадлежат разным поверхностным фрагментам.

*1) Площадь сферического треугольника.(Рис. 4)*

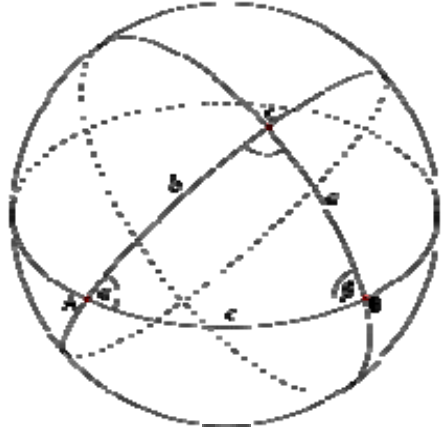

**Рис. 4** Сферический треугольник

Сумма s углов сферического треугольника α ,β, γ :

$s = \alpha + \beta + \gamma$

( 12 )

всегда меньше 3π и больше π. Величина

$\varepsilon = s - \pi$

( 13 )

называется сферическим избытком. Площадь сферического треугольника S определяется по формуле Жирара:

$S = R^2 \varepsilon$,

( 14 )

где  *R*-  радиус сферы.

   *2) Площадь тороидального треугольника.(Рис. 5)*

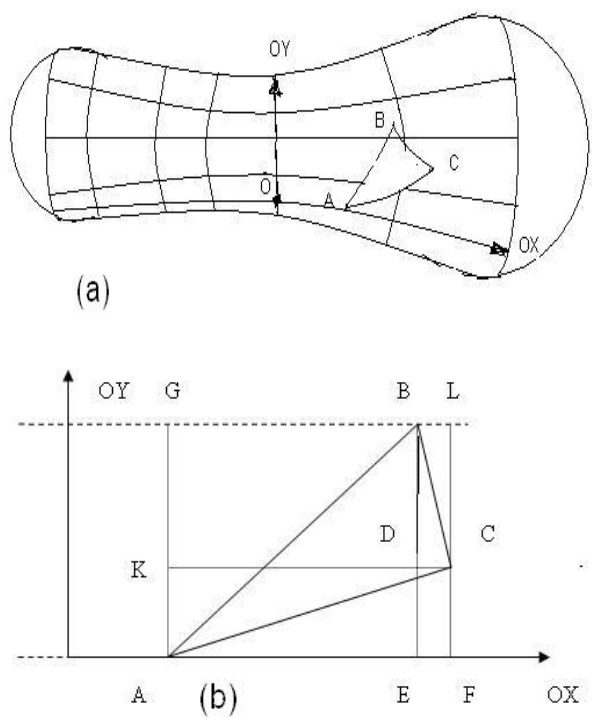

**Рис. 5** (a)Тороидальный треугольник ABC (b) Его отображение на плоскость, пропорции искажены

Рассмотрим тороидальный треугольник ABC на торе.

Определим криволинейную систему коорлинат, образованной образующими тора двух типов:

OX, OY –оси, идущие вдоль образующих (OY – круги, перпендекулярные оси тора, OX – точки касания между тором и сферой обкатки)

При этом возьмем некие три точки A, B, C. Базисные оси, идущие вдоль образующих OX, OY могут всегда быть выбраны следующим образом:

1) Точка C лежит между точками A и B по оси ОУ.
2) Положитеельное направление оси OX от отрезка [AB] к точке C
3) Для точек $A(x_A, y_A)$ и $B(x_B, y_B)$ $x_B > x_A$, $y_B > y_A$
4) Линии, соединяющие пары точек AC, AB, CB мы можем выбрать не всегда совпадающими с геодезическими тора. Выберем их как отрезки

линий между точками, которые являются образом геодезических линий цилиндра, получающимся отоборажением тора на этот цилиндр. Это отображение делается так : образующие тора ОХ отображаются на образующие цилиндра, параллельные оси цилиндра, с сохранением длины, образующие тора ОУ отображаются на образующие цилиндра, перпендекулярные оси цилиндра, с сохранением угловых размеров.

5) $\alpha_A$ - угловое положение вдоль образующей ОХ   $x_A = R_{ox}\alpha_A$, $R_{OX}$ - радиус сферы обкатки тора, h - расстояние от центра сферы обкатки до прямой, соединяющей опорные сферы тора.

6) $\varphi_{AB}$ - угловое расстояние между точками А и В вдоль образующей OY

Тогда площадь тороидального треугольников и четырехугольников:

Площадь квадрата EAGB

$$S_{EAGB} = \varphi_{AB} R_{OX} (h\alpha - R_{OX} \sin(\alpha))\vert_{\alpha_A}^{\alpha_B}.$$

( 15 )

Площадь треугольников  ABE, BCD и ACF

$$S_{ABE} = \varphi_{AB} \frac{R_{OX}}{\alpha_B - \alpha_A} (h\frac{(\alpha - \alpha_A)^2}{2} - R_{OX}((\alpha - \alpha_A)\sin(\alpha) + \cos(\alpha)))\vert_{\alpha_A}^{\alpha_B},$$  ( 16 )

$$S_{BCD} = \varphi_{BC} \frac{R_{OX}}{\alpha_B - \alpha_C} (h\frac{(\alpha - \alpha_C)^2}{2} - R_{OX}((\alpha - \alpha_C)\sin(\alpha) + \cos(\alpha)))\vert_{\alpha_B}^{\alpha_C},$$  ( 17 )

$$S_{ACF} = \varphi_{AC} \frac{R_{OX}}{\alpha_A - \alpha_C} (h\frac{(\alpha - \alpha_C)^2}{2} - R_{OX}((\alpha - \alpha_C)\sin(\alpha) + \cos(\alpha)))\vert_{\alpha_A}^{\alpha_C}.$$  ( 18 )

Площадь треугольника  ABC

$$S_{ABC} = S_{ABE} - S_{ACF} + sign(x_C - x_B)(S_{EDCF} + S_{BCD}).$$  ( 19 )

### 3) Площади тругольников, чьи вершины не принадлежат одному и тому же поверхностному фрагменту.

Треугольник образован вершинами $r_1, r_2, r_3$. Используем для расчета их площади S формулу площади плоского треугольника.

Стороны треугольника:

$$a = r_2 - r_1 \quad b = r_3 - r_1. \quad (20)$$

Площадь треугольника S:

$$S = \frac{1}{2}|[a \times b]|.$$

( 21 )

### 2.5.5.2 Формирование многоугольных поверхностныч элементов и определение их параметров

На основе треугольных поверхностных элементов формируются *многоугольные* поверхностные элементы. Центры многоугольных элементов совпадают с вершинами треугольных элементов. Вершины многоугольных элементов – центры сторон треугольных элементов, выходящие из центров многоугольных элементов и центры тяжести (точка пересечения медиан) этих треугольных элементов.

Площадь многоугольного элемента $S^M$ - сумма третей от площедей треугольных элементов $S^{tr}_i$ с вершиной в центре многоугольного элемента. Для каждого узла сетки триангуляции производиться суммирование площадей тех треугольников, для которых данный узел является вершиной. Полученная суммарная площадь делиться на три, полученный результат запоминается как площадь поверхностного элемента с координатами и нормалью соответствующих данному узлу сетки триангуляции

$$S^M = \frac{\sum_i S_i^{tr}}{3}.$$

( 22 )

### 2.5.5.3 Поверхностные элементы на SAS

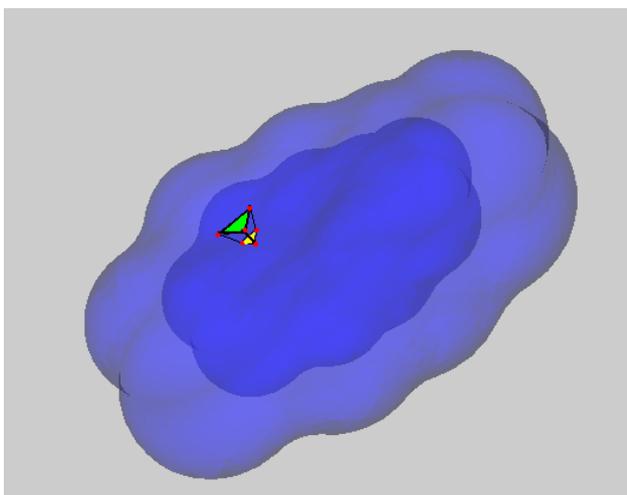

**Рис. 6**. Преобразование поверхности типа SES в поверхность типа SAS

Пусть имеется поверхность после первичной обкатки.

Пусть имеется треугольник на поверхности SES (*r₁*,*r₂*,*r₃*)

Образ этого треугольника на SAS (*r₁ₙ*,*r₂ₙ*,*r₃ₙ*) (Рис.6)

$r_{1n} = r_1 + n_1 R_{pr},$

( 23 )

$r_{2n} = r_2 + n_2 R_{pr},$

( 24 )

$r_{3n} = r_3 + n_3 R_{pr},$ \hfill ( 25 )

где $R_{pr}$ – радиус первичной обкатки, *n₁*, *n₂*, *n₃*-нормали в соответствущий точках.

Тогда

1) поверхностный элемент на первичной сфере, опирающейся на три атома, отображается в точку. Его площадь на SAS нулевая
2) поверхностный элемент на торе, опирающимся на два атома, отображается в линию. Его площадь на SAS нулевая
3) поверхностный элемент на атоме отображается в сферический тругольник. Его площадь на SAS пропорцианальна его площади на SES: $S_{ses}((R_{atom} + R_{pr})/R_{atom})^2$
4) граничный поверхностный элемент между тором и первичной сферой, опирающейся на три атома – нулевая площадь
5) граничный треугольный поверхностный элемент, опирающийся на первичный тор и атом, первичную сферу и атом или первичные тор, сферу и атом – площадь S считается по формуле для трех точек $(\boldsymbol{r}_{1n}, \boldsymbol{r}_{2n}, \boldsymbol{r}_{3n})$ :

$$\boldsymbol{a} = \boldsymbol{r}_{2n} - \boldsymbol{r}_{1n}, \tag{26}$$

$$\boldsymbol{b} = \boldsymbol{r}_{3n} - \boldsymbol{r}_{1n}, \tag{27}$$

$$S = \frac{|\boldsymbol{a} \times \boldsymbol{b}|}{2}. \tag{28}$$

### 2.5.5.4 Вычисление площади и объема поверхностей SES и SAS

Вычисление площади поверхности производиться суммированием площадей всех поверхностных элементов

$$S = \oint_S dS .$$

( 29 )

Вычисление объема поверхности производиться по следующей формуле:

$$V = \frac{1}{3}\oint_S (\boldsymbol{r}\cdot\boldsymbol{n})dS .$$

( 30 )

***r*** – радиус-вектор текущей точки поверхности, ***n*** - ее нормаль

## 3. Выводы.

Представленный алгоритм и основанная на нем программа позволяют быстро и надежно строить триангуляцию гладкой поверхности. Поверхность получается раскрашенной в зависимости от типа ближайших атомов, что удобно для визуализации, и триангулированной, что позволяет не только вычислять ее поверхность и ограниченный ею объем, но и решать с хорошей точностью интегральные уравнения, используемые в континуальной модели растворителя. Полученная с помощью данной программы поверхность может применяться как для целей визуализации молекулы, это особенно актуально для больших белковых молекул, так и для целей вычисления сольватационных вкладов в энергию взаимодействия молекул друг с другом при наличии внешней среды.

Алгоритм триангуляции может быть легко обобщен на случай любой гладкой поверхности, в том числе гладкой поверхности уровня.

**Благодарности.**

Данная работа является более детальным изложением и дальнейшим развитием статьи [35]. Авторы выражают глубокую признательность всем авторам этой публикации за проделанную работу, которая послужила основой и для данной статьи.

Список литературы